\documentclass[
preprint,
 amsmath,amssymb,
 aps,
]{revtex4-1}

\usepackage{graphicx}
\usepackage{dcolumn}
\usepackage{bm}

\newtheorem{theorem}{Theorem}
\newtheorem{corollary}{Corollary}

\begin{document}


\title{Rainich Conditions in (2+1)-Dimensional Gravity}

\author{D. S. Krongos and C. G. Torre}
\affiliation{%
 Department of Physics, Utah State University\\
 Logan, UT 84322-4415, USA
}%

\date{November 7, 2016}

\begin{abstract}
In (3 + 1) spacetime dimensions the Rainich conditions are a set of equations expressed solely in terms of the metric tensor which are equivalent to the Einstein-Maxwell equations for non-null electromagnetic fields. Here we provide the analogous conditions for (2 + 1)-dimensional gravity coupled to electromagnetism. Both the non-null and null cases are treated. The construction of these conditions is based upon reducing the problem to that of gravity coupled to a scalar field, which we have treated elsewhere.  These conditions can be easily extended to other theories of (2 + 1)-dimensional gravity.  For example, we apply the geometrization conditions to topologically massive gravity coupled to the electromagnetic field and obtain a family of plane-fronted wave solutions.
\end{abstract}

\pacs{Valid PACS appear here}

\clearpage
\maketitle
\thispagestyle{empty}


\section{\label{sec:level1}Introduction}
\setcounter{page}{1}

The  Rainich geometrization of the electromagnetic field replaces the Einstein-Maxwell equations with an equivalent set of equations which are (1) expressed solely in terms of the spacetime metric, and (2) provide a method by which the electromagnetic field  can be reconstructed from the geometry of the spacetime. The case originally explored by Rainich \cite{Rainich}, and later by Misner and Wheeler \cite{MW}, was the geometrization of the electromagnetic field in four spacetime dimensions. The geometrization was brought to completion only for non-null fields, and took advantage of special features of four dimensions, {\it e.g.,} the Hodge dual of a 2-form is again a 2-form.   The work of Rainich, Misner and Wheeler has since been extended to include the null case and to include various matter fields besides the electromagnetic field; see \cite{KrongosTorre2015} and references therein. To our knowledge, for the electromagnetic field in dimensions other than four only partial geometrization results have been obtained, limited to the ``algebraic'' part of the Rainich conditions \cite{Senovilla}.  As we shall show, in (2+1) dimensions it is possible to complete the geometrization process for the electromagnetic field for   non-null fields and for null fields \cite{Deser1}. The simplifying feature of  three spacetime dimensions is that the problem of geometrization of an electromagnetic field can be reduced to that of a scalar field, whose solution is known \cite{Kuchar, KrongosTorre2015}. 

We will state the geometrization conditions in the next section.  Section III will explain how these results are proved.  Section IV illustrates the results of \S II by  using the geometrization conditions to obtain the static, charged BTZ black hole.  Section V explains how to apply the geometrization results developed here to more general theores of (2+1) gravity.  In particular, we use the results of \S II to obtain a family of  pp-wave solutions to topologically massive gravity coupled to the electromagnetic field. 

\section{\label{Results} Geometrization of the electromagnetic field in (2+1) dimensions}

Let $(M, g)$ be a (2 + 1)-dimensional spacetime with signature $(-++)$. The Einstein-Maxwell equations with electromagnetic 2-form $F$ and cosmological constant $\Lambda$ are given by Einstein's equations
\begin{equation}
	G_{ab} + \Lambda g_{ab} = q \left(F_{ac} F_{b}^{\;c} - \frac{1}{4} g_{ab} F_{de} F^{de} \right)
	\label{einstein_em}
\end{equation}
along with the source-free Maxwell equations
\begin{equation}
	F_{ab\,;}{}^{a} = 0,\quad F_{[ab\,;\,c]} = 0.
	\label{div_e}
\end{equation}
Here a semi-colon denotes covariant differentiation with respect to the Christoffel connection,  $G_{ab}$ is the Einstein tensor, and $q>0$ represents Newton's constant.  All fields on $M$ will be assumed to be smooth.

We note that  the Einstein-Maxwell equations admit a discrete symmetry: if $(g, F)$ is a solution to these equations then so is $(g, -F)$.   For this reason the electromagnetic field can be recovered from the geometry only up to a sign. 

We say an electromagnetic field is {\it null} in a given region $U\subset M$ if $F_{ab}F^{ab} = 0$ on $U$, and the field is {\it non-null} in a region $U\subset M$ if $F_{ab}F^{ab} \neq 0$ on $U$.  As in (3+1) dimensions, the null and non-null cases must be treated separately. 

Define
\begin{equation}
G = G_a^a,\quad _2G = G^a_bG^b_a,\quad _3G=G^a_bG^b_cG^c_a.
\end{equation}
The (2+1)-dimensional version of the Rainich geometrization of non-null electromagnetic fields is as follows. 

\begin{theorem} Let $(M, g)$ be a  (2+1)-dimensional spacetime. The following  are necessary and sufficient conditions on $g$ such that on $U\subset M$ there exists a non-null electromagnetic field $F$ with $(g, F)$ being a solution of the Einstein-Maxwell equations (\ref{einstein_em})--(\ref{div_e}):
\begin{equation}
	_2G - \frac{1}{3} G^2 \neq 0,
	\label{condition_1}
\end{equation}
\begin{equation}
	H_{ab} w^a w^b > 0, \;\;\;\; \text{ for some } w^a,
	\label{condition_5}
\end{equation}
\begin{equation}
	B= \Lambda,
	\label{condition_2}
\end{equation}
\begin{equation}
	H_{a[b} H_{c]d} = 0,
	\label{condition_3}
\end{equation}
\begin{equation}
	H_{ab} H_{c[d;e]} + H_{ac} H_{b[d;e]} + H_{bc;[d} H_{e]a} = 0,
	\label{condition_4}
\end{equation}

where 
\begin{equation}
	B = \frac{1}{2} \frac{ \frac{1}{3} G\; _2G - {}_3G}{_2G - \frac{1}{3}G^2},
	\label{B_def}
\end{equation}
and
\begin{equation}
	H_{ab} = G_{ab} - (G + 2B)g_{ab}.
	\label{H_def}
\end{equation}
These conditions hold everywhere on $U$.

\label{nonnulthm}
\end{theorem}

\begin{corollary}
Let a metric $g$ satisfy the conditions of Theorem \ref{nonnulthm}. Then $(g, F)$ satisfy the Einstein-Maxwell equations on $U$ with the non-null electromagnetic field $F$ determined up to a sign from\begin{equation}
F_{ab} = \epsilon_{abc} v^c, \quad v_a v_b = \frac{1}{q}H_{ab}.
\end{equation} 
\label{c1}
\end{corollary}

The geometrization in the null case is as follows. 

\begin{theorem}
Let $(M, g)$ be a  (2+1)-dimensional spacetime. The following are necessary and sufficient conditions on $g$ such that on $U\subset M$ there exists a null electromagnetic field $F$ with $(g, F)$ being a solution of the Einstein-Maxwell equations (\ref{einstein_em})--(\ref{div_e}):
\begin{equation}
G = -3\Lambda,
\label{nsf0}
\end{equation}
\begin{equation}
	S_{ab} w^a w^b > 0 \quad{\rm for\ some\ } w^a,
\label{nsf4}
\end{equation}
\begin{equation}
	S_{a[b}S_{c]d} = 0,
\label{nsf1}
\end{equation}
\begin{equation}
	 S_{ab}S_{c[d;e]} + S_{ac}S_{b[d;e]} + S_{bc;[d} S_{e]a} = 0, 
\label{nsf2}
\end{equation}
where $S_{ab}= G_{ab} - \frac{1}{3}G_c^c\, g_{ab}$ is the trace-free Einstein (or Ricci) tensor. These conditions hold everywhere on $U$.

\label{nullthm}
\end{theorem}

\begin{corollary}
Let a metric $g$ satisfy the conditions of Theorem \ref{nullthm}.  Then $(g, F)$ satisfy the Einstein-Maxwell  equations on $U$ with a null electromagnetic field $F$ determined up to a sign from
\begin{equation}
F_{ab} = \epsilon_{abc} v^c,\quad v_a v_b= \frac{1}{q}S_{ab}.
\end{equation}
\end{corollary}

As also happens  in (3+1) dimensions, for both the null and non-null cases the Rainich conditions split into conditions which are algebraic in the Einstein (or Ricci) tensor and conditions which involve derivatives of the Einstein tensor.     In (3+1) dimensions the non-null Rainich conditions involve up to 4 derivatives of the metric, while the null Rainich conditions can involve as many as 5 derivatives \cite{Torre2014}.  From the above theorems, in (2+1) dimensions both the null and non-null conditions involve up to 3 derivatives of the metric. 

\section{\label{sec:level2}Proofs}

We now prove the results stated in the previous section. The electromagnetic field $F$ is a two-form in three spacetime dimensions, so (at least locally) we can express it as the Hodge dual of a one-form $v$,
\begin{equation}
	F_{ab} = \epsilon_{ab}^{\;\;\;c}v_c,\quad v_a = -\frac{1}{2}\epsilon_{abc} F^{bc},
	\label{F_dual_v}
\end{equation}
where $\epsilon_{abc}$ is the volume form defined by the Lorentz signature metric, and which satisfies
\begin{equation}
	\epsilon^{abc}\epsilon_{def} = -3! \delta^{[a}_d\delta^b_e\delta^{c]}_f.
\end{equation}
The Einstein-Maxwell equations (\ref{einstein_em})--(\ref{div_e}) can then be rewritten as
\begin{equation}
	G_{ab} + \Lambda g_{ab} = q \left(v_a v_b - \frac{1}{2} g_{ab} v_c v^c \right),
	\label{einstein_em_v}
\end{equation}
\begin{equation}
v_{[a;b]} = 0 = v^a_{;a}.
\end{equation}
These equations are locally equivalent to gravity coupled to a scalar field, where the scalar field $\phi$ is massless and minimally-coupled. The correspondence is  via $v_a = \nabla_a \phi$.   Consequently, the geometrization runs along the same lines as the scalar field case, found  in \cite{KrongosTorre2015}. 

We begin with Theorem 1.  To see that the conditions are necessary, we consider a metric $g$ and non-null electromagnetic field $F$ satisfying the Einstein-Maxwell equations.  From (\ref{einstein_em_v}) it follows that
\begin{equation}
	_2G - \frac{1}{3} G^2 = \frac{2}{3}q^2 \left(v_c v^c \right)^2 \neq 0,
\end{equation}
\begin{equation}
B = \Lambda,\quad H_{ab} = q v_a v_b,
\end{equation}
and
\begin{equation}
	H_{ab}H_{c[d;e]} + H_{ac}H_{b[d;e]} + H_{bc;[d}H_{e]a} = 2q^2v_a v_b v_c v_{[d;e]}   = 0,
\end{equation}
from which it follows that the conditions (\ref{condition_1})--(\ref{condition_4}) in Theorem 1 are necessary.

Conversely, suppose equations (\ref{condition_1})--(\ref{condition_4}) are satisfied. From Eqs. (\ref{condition_5}) and (\ref{condition_3}) there exists a one-form $v_a$ such that 
\begin{equation}
	H_{ab} = qv_a v_b.
\label{Hv}
\end{equation}
(See  \cite{KrongosTorre2015} for a proof.)   Equation (\ref{condition_1}) implies $v_a v^a \neq 0$.  
Equation (\ref{condition_4}) becomes $2v_av_bv_cv_{[d;e]} = 0$, so that  $v_{[a;b]} = 0$.
Taking account of condition (\ref{condition_2}), we now have 
\begin{align}
	G_{ab} =  &q \left(v_a v_b - \frac{1}{2} g_{ab} v_c v^c \right) - \Lambda g_{ab}, \\ 
	&v_{[a;b]} = 0,\  v_av^a\neq0.
\end{align}
From the contracted Bianchi identity, $\nabla^b G_{ab} = 0$, we get
\begin{equation}
	v_b  v^a_{;a} = 0,
\end{equation}
so that the Einstein-Maxwell equations are satisfied.  The construction of the electromagnetic field from the metric described in Corollary \ref{c1} follows from solving the algebraic relations (\ref{Hv}) for $v_a$ and then using (\ref{F_dual_v}).

The null case, described in Theorem 2 and Corollary 2, is established as follows.   As before, begin by assuming the Einstein-Maxwell equations are satisfied in the null case, that is, with $v_av^a = 0$.   The trace and trace-free parts of the Einstein equations yield, respectively,
\begin{equation}
G = -3\Lambda,\quad S_{ab} = qv_a v_b,
\label{eqnull}
\end{equation}
These equations and the Maxwell equations $v_{[a;b]} = 0$ imply the necessity of the conditions listed in Theorem 2.  Conversely, granted the conditions of Theorem 2, it follows in a similar fashion as in the proof of Theorem 1 that equations (\ref{eqnull}) hold with $v_av^a=0$, and that the Maxwell equations $v_{[a;b]} = 0$ are satisfied.  The contracted Bianchi identity again implies $v^a_{;a}=0$.  The construction of the electromagnetic field from the metric described in Corollary 2 follows from solving the algebraic relations (\ref{eqnull}) for $v_a$ and then using (\ref{F_dual_v}).

\section{\label{BTZ}Example: BTZ Black Hole}
As an illustration of these geometrization conditions we investigate static, rotationally symmetric solutions to the Einstein-Maxwell equations. Begin with the following ansatz for the metric:
\begin{equation}
	g = -f(r)\, dt\otimes dt + \frac{1}{f(r)}\, dr \otimes dr + r^2 d\theta \otimes d\theta,
\label{SSSmetric}
\end{equation}
where $f(r)$ is to be determined by the geometrization conditions. 
The algebraic condition (\ref{nsf1}) from Theorem 2 would imply the metric (\ref{SSSmetric}) is Einstein, so there can be no electromagnetic field in the null case.    In the non-null case the conditions of Theorem 1 reduce to a remarkably simple  linear third-order differential equation
\begin{equation}
	f'''(r) + \frac{1}{r} f''(r) - \frac{1}{r^2} f'(r) = 0,
\end{equation}
which has the solution 
\begin{equation}
	f(r) = c_1 + c_2 \ln{r} + c_3 r^2,
\label{fsol}
\end{equation} 
where $c_1, c_2$, and $c_3$ are constants of integration.  Eq. (\ref{condition_5}) requires $c_2 < 0$. 
From equation (\ref{condition_2}) the form of $f(r)$ given in (\ref{fsol}) corresponds to a cosmological constant
\begin{equation}
	\Lambda = -c_3,
\end{equation}
and, from Corollary \ref{c1}, to an electromagnetic field
\begin{equation}
	F = \pm\frac{\sqrt{-c_2}}{r} dt \wedge dr.
\end{equation}
With the identifications 
\begin{equation}
c_1 = -M, \quad c_2 = -\frac{1}{2}Q^2, \quad c_3 = \frac{1}{\ell^2}
\end{equation} 
we obtain the static charged BTZ solution \cite{BTZ}.

\medskip

\section{\label{Extension}Extension to other metric theories of gravity}
The geometrization conditions obtained here can be extended to other metric theories of (2 + 1)-dimensional gravity coupled to electromagnetism provided the action functional $S$ for the system takes the form 
\begin{equation}
S = S_1[g] -\frac{2}{q} \Lambda V[g]+ S_2[g, F],
\end{equation}
where $S_1$ is diffeomorphism invariant,   $S_2$ is the usual  action for the electromagnetic field on a three-dimensional spacetime with metric $g$, and $V$ is the volume functional.   In the field equations, theorems, and corollaries given above one simply makes the replacement 
\begin{equation}
G^{ab}\longrightarrow E^{ab} =  - \frac{1}{\sqrt{|g|}} \frac{\delta S_1}{\delta g_{ab}}.
\end{equation}
The  identity $E^{ab}{}_{;b}=0$ still holds because of the diffeomorphism invariance of $S_1$; all the proofs remain unchanged. 

As a simple application of this result, we suppose  the action $S_1$ is a linear combination of the Einstein-Hilbert action and the Chern-Simons action constructed from the metric-compatible connection.  The field equations are the Maxwell equations (\ref{div_e}) along with
\begin{equation}
\alpha G_{ab} + \beta Y_{ab} + \Lambda g_{ab} = q \left(F_{ac} F_{b}^{\;c} - \frac{1}{4} g_{ab} F_{de} F^{de} \right),
\end{equation}
 where $Y_{ab}$ is the Cotton-York tensor \cite{CottonYork} and  $\alpha$, $\beta$ are constants.  These are the equations of topologically massive gravity \cite{Deser2} coupled to the electromagnetic field.  We ask whether there are any solutions of the pp-wave type, admitting a covariantly constant null vector field.  Using the usual metric ansatz
\begin{equation}
g = -2 du \odot dv + dx\otimes dx + f(u, x)\, du \otimes du,
\label{ppwave}
\end{equation}
it follows that condition (\ref{condition_1}) in Theorem 1 is not satisfied, so only null solutions are possible.  For this metric $E^a_a=0$;  (\ref{nsf0}) then implies we can only get a solution for $\Lambda = 0$.  The   conditions (\ref{nsf4}) -- (\ref{nsf2}) of Theorem 2 reduce to 
\begin{equation}
\alpha \frac{\partial^3f}{\partial x^3} + \beta \frac{\partial^4f}{\partial x^4} = 0,
\end{equation}
with solution (assuming $\beta\neq0$)
\begin{equation}
f(u, x) = a_0(u)+a_1(u)x  - a_2(u)x^2+b(u)e^{-\frac{\alpha}{\beta}x},
\label{ppsol}
\end{equation}
where  $\alpha\, a_2(u) > 0$, and $a_0(u), a_1(u), a_2(u), b(u)$ are otherwise arbitrary.  From Corollary 2 the electromagnetic field is given by
\begin{equation}
F = \sqrt{\alpha a_2(u) /q}\,  du \wedge dx.
\label{ppEM}
\end{equation}
Evidently, the term in $f(u, x)$ quadratic in $x$ determines (or is determined by) the electromagnetic field.  The York tensor vanishes, {\it i.e.}, the metric is conformally flat, if and only if $b(u)=0$.

%
\noindent
\begin{acknowledgments}
The computations in \S \ref{BTZ}--\ref{Extension},  were performed using the {\it DifferentialGeometry} software package \cite{DG}.   This work was supported in part by National Science Foundation grants  ACI-1642404, OCI-1148331 to Utah State University.
\end{acknowledgments}

\end{document}